\documentclass[sigconf, nonacm]{acmart}

\newcommand\vldbdoi{XX.XX/XXX.XX}
\newcommand\vldbpages{XXX-XXX}
\newcommand\vldbvolume{14}
\newcommand\vldbissue{1}
\newcommand\vldbyear{2020}
\newcommand\vldbauthors{\authors}
\newcommand\vldbtitle{\shorttitle}
\newcommand\vldbavailabilityurl{https://github.com/Flitternie/EnumGRPO}
\newcommand\vldbpagestyle{plain}

\newcommand{\eg}{\emph{e.g.,} }

\usepackage{xspace}
\newcommand{\name}{agentic query execution\xspace}
\definecolor{revisionblue}{RGB}{0,70,160}
\newcommand{\revise}[1]{{\color{black}#1}}

\usepackage{algorithm}
\usepackage{algpseudocode}
\usepackage{enumitem}
\usepackage[framemethod=TikZ]{mdframed}
\usepackage{makecell}
\usepackage{colortbl}
\usepackage{listings}
\usepackage{titlesec}
\titleformat{\paragraph}[runin]{\normalfont\bfseries}{}{0pt}{}[]
\titlespacing{\paragraph}{0pt}{0.5ex plus 0.2ex minus 0.1ex}{0.5em}

\usepackage{cleveref}

\mdfdefinestyle{exbar}{%
  topline=false, bottomline=false, rightline=false,
  linewidth=1.5pt, linecolor=black!25,
  innerleftmargin=0.7em, innerrightmargin=0pt,
  innertopmargin=0.2em, innerbottommargin=0.2em,
  leftmargin=0.3em, skipabove=1.5ex, skipbelow=0.5ex}

{\theoremstyle{definition}\newtheorem{example}{Example}}

\begin{document}
\title{Cost-Aware Optimization for Agentic Query Execution}

\author{Lunyiu Nie}
\affiliation{%
  \institution{The University of Texas at Austin}
}
\email{lynie@utexas.edu}

\author{Yilin Xia}
\affiliation{%
  \institution{University of Illinois Urbana-Champaign}
}
\email{yilinx2@illinois.edu}

\author{Yiren Liu}
\affiliation{%
  \institution{University of Illinois Urbana-Champaign}
}
\email{yirenl2@illinois.edu}

\author{Christopher Jermaine}
\affiliation{%
  \institution{Rice University}
}
\email{cmj4@rice.edu}

\author{Swarat Chaudhuri}
\affiliation{%
  \institution{The University of Texas at Austin}
}
\email{swarat@cs.utexas.edu}

\begin{abstract}
    Classical query optimization searches over algebraically equivalent plans that differ only in cost. This assumption breaks once LLM-backed operators enter the picture: their placement, ordering, and granularity jointly determine both cost and answer quality, and the right choice among the alternatives is often revealed only at runtime. We formalize this setting as \textbf{\name}, a paradigm in which agent-based planning is interleaved with execution, and \textbf{agent workflow optimization} becomes the analogue of classical query optimization. We then present \textbf{EnumGRPO}, a self-improving optimizer for this setting. During a learning stage, EnumGRPO enumerates query plans over decisions such as execution paradigm, operator type, operator placement, selectivity scope, and projection width, then distills quality-cost feedback into reusable planning heuristics via in-context reinforcement learning. \revise{On SWAN, EnumGRPO achieves higher execution accuracy than LOTUS, Palimpzest, and Agentic BlendSQL, while cutting LLM-operator cost by up to 317$\times$. Without retraining, the learned experiences transfer to another LLM family, SemBench workload, and purely relational Spider and BIRD workloads, improving or preserving quality at higher cost efficiency.}
\end{abstract}

\maketitle

\pagestyle{\vldbpagestyle}
\begingroup\small\noindent\raggedright\textbf{PVLDB Reference Format:}\\
\vldbauthors. \vldbtitle. PVLDB, \vldbvolume(\vldbissue): \vldbpages, \vldbyear.\\
\href{https://doi.org/\vldbdoi}{doi:\vldbdoi}
\endgroup
\begingroup
\renewcommand\thefootnote{}\footnote{\noindent
This work is licensed under the Creative Commons BY-NC-ND 4.0 International License. Visit \url{https://creativecommons.org/licenses/by-nc-nd/4.0/} to view a copy of this license. For any use beyond those covered by this license, obtain permission by emailing \href{mailto:info@vldb.org}{info@vldb.org}. Copyright is held by the owner/author(s). Publication rights licensed to the VLDB Endowment. \\
\raggedright Proceedings of the VLDB Endowment, Vol. \vldbvolume, No. \vldbissue\ %
ISSN 2150-8097. \\
\href{https://doi.org/\vldbdoi}{doi:\vldbdoi} \\
}\addtocounter{footnote}{-1}\endgroup

\ifdefempty{\vldbavailabilityurl}{}{
\vspace{.3cm}
\begingroup\small\noindent\raggedright\textbf{PVLDB Artifact Availability:}\\
The source code, data, and/or other artifacts have been made available at \url{\vldbavailabilityurl}.
\endgroup
}
\section{Introduction}
\label{sec:introduction}

Classically, query optimization applies algebraic transformations without changing the answer by design: one can push a selection past a join, reorder associative operators, or choose between index scans and hash joins, and the result is guaranteed to be equivalent. The optimizer's job is to navigate the space of these equivalent plans and pick the most efficient one \cite{selinger1979,chaudhuri1998overview,graefe1995cascades}.

This narrow view, that all candidate plans produce the same answer and differ only in cost, breaks down once Large Language Model (LLM)-backed operators enter the picture. A recent line of hybrid database systems \cite{blendsql2024,zhao2025hybrid,palimpzest2025,weaver2025,patel2025semanticoptimization,kumarasinghe2026ipdb} has extended SQL with semantic operators that invoke LLMs to normalize values, resolve ambiguities, and fill knowledge gaps that the schema alone cannot supply.
Unlike a hash join or an index scan, however, an LLM operator is \emph{approximate}: it can produce different answers depending on how much context it receives, which rows it processes, and where in the plan it is placed. At the same time, each LLM invocation is orders of magnitude more expensive than a relational operator.
The result is that structural decisions about the workflow, the very choices that a classical optimizer treats as cost-neutral equivalences, now jointly determine both answer quality and dollar cost. Finding the best plan is no longer a search over equivalent alternatives. It is a search over plans that differ in both cost \emph{and} correctness.

\begin{figure}[!t]
  \centering
  \includegraphics[width=\columnwidth]{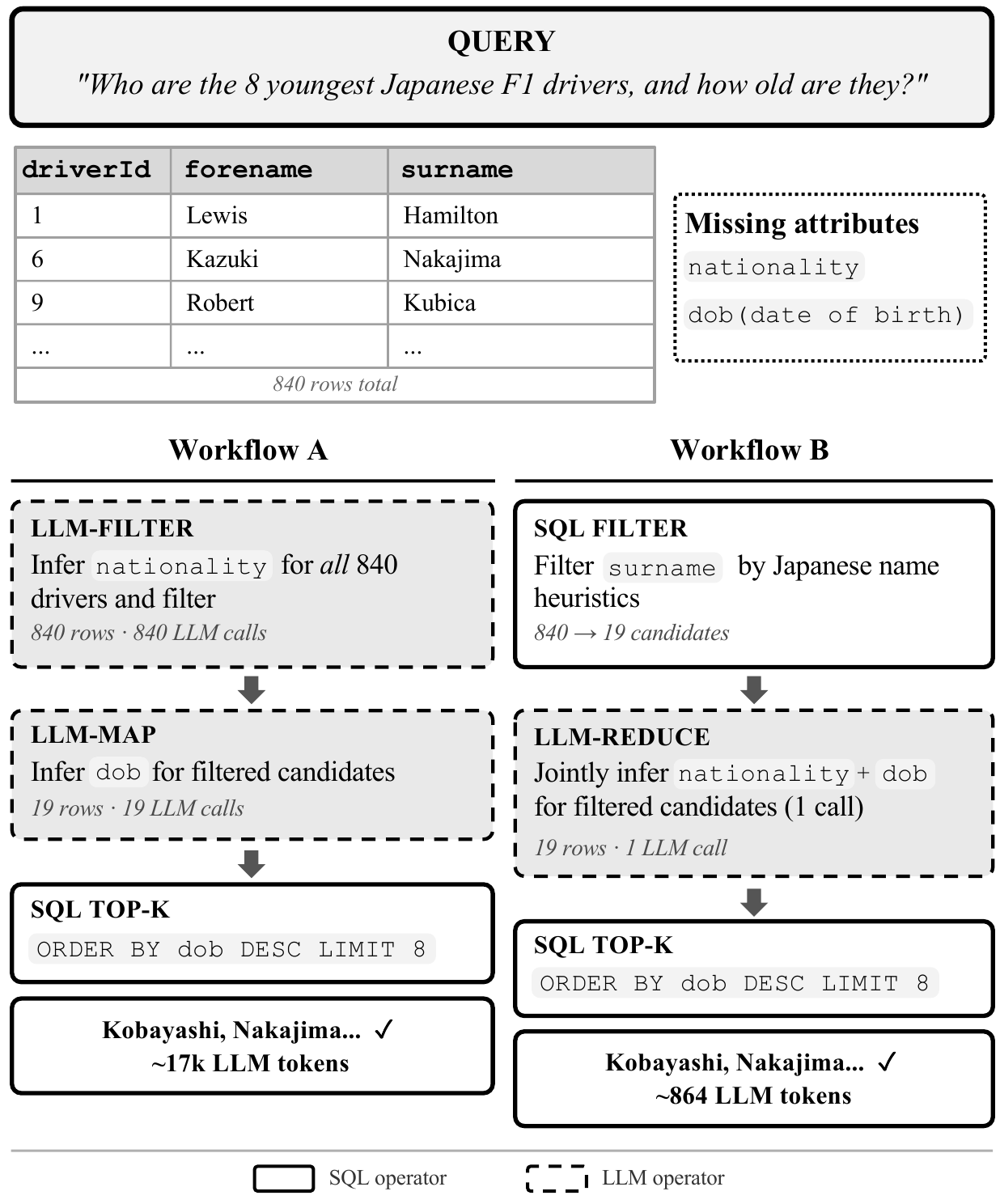}
  \caption{Two valid workflows for the same query. \texttt{nationality} and \texttt{dob} are absent from the table. Both return the correct answer, but Workflow~B narrows candidates in SQL and batches LLM inference, reducing token cost by $20{\times}$.}
  \label{fig:motivation}
\end{figure}

Figure~\ref{fig:motivation} illustrates this tradeoff for the query ``Who are the 8 youngest Japanese F1 drivers, and how old are they?'' The table lacks \textit{nationality} and \textit{date-of-birth}, which an LLM can supply. Workflow A invokes the LLM for every driver to infer nationality, filters the result, and invokes it again for birth dates. Workflow B first narrows candidates with SQL and then batches the survivors. Both happen to return the correct answer, although this cannot be known from the query alone, while Workflow B costs $20{\times}$ less. How should an optimizer jointly manage cost and accuracy in such pipelines?

This requires an optimizer whose decisions are made as part of query execution. Existing hybrid systems \cite{blendsql2024, DBLP:journals/corr/abs-2407-11418,palimpzest2025,patel2025semanticoptimization,kumarasinghe2026ipdb} separate optimization from execution: the plan is fixed before any tuple is processed and never revised against intermediate results. But for many analytical workloads, the need for semantic reasoning emerges from query execution itself. The decisive choices depend on cardinalities, value distributions, and schema and data semantics that only become observable at runtime. Intermediate results determine whether to use an LLM operator and, if so, its appropriate form, plan position, degree of selectivity, and context size.

The unit of execution is therefore no longer a pre-specified hybrid query but an \emph{agent workflow} that we call \textbf{\name}. Rather than committing to a fully specified SQL+LLM plan, the system executes one step, observes the resulting relation, and decides the next action: which operator to invoke, on which intermediate relation, and at what granularity. SQL remains responsible for deterministic relational computation, while LLM operators are invoked only when intermediate results indicate that semantic reasoning is required. We use \name as the execution setting, but the core systems problem is \textbf{agent workflow optimization}: better runtime decision-making over dynamically constructed workflows. The optimizer must decide how to decompose a task into steps, when to push narrowing operations into SQL before invoking an LLM, and how much data each LLM call should inspect, while taking both accuracy and cost into account. This connects our setting to adaptive query processing and learned query optimization \cite{avnur2000eddies,bao2021,balsa2022,lero2023,skinnerdb2021}.

To address this, we introduce \textbf{EnumGRPO}, a self-improving optimizer for agent workflow optimization. During a learning stage, EnumGRPO explicitly enumerates diverse candidate workflows for each query over \revise{five planning axes with conditional applicability} (execution paradigm, operator type, operator placement, selectivity scope, and projection width). Each axis has a classical analogue, such as predicate pushdown, late materialization, or access-path selection~\cite{selinger1979,abadi2008column}, but now governs both cost and answer quality, not cost alone. Candidates are executed and scored by a joint quality-cost objective. Group-relative advantages isolate the effect of structural decisions from query-level difficulty, and a four-stage experience distillation pipeline converts contrastive feedback into reusable planning heuristics. At query time, the learned policy applies these cross-query heuristics without runtime enumeration to navigate the quality-cost Pareto frontier of agentic workflows.

The entire procedure operates through \emph{in-context reinforcement learning}: policy improvement flows through the agent's prompt, without gradient updates to model weights. We adopt this design for two reasons. First, the learned heuristics are encoded as natural language and remain human-readable and auditable. Second, the optimizer is portable: because the planning heuristics live in the prompt rather than the model weights, it is decoupled from any specific LLM backbone. This is essential for a long-lived DBMS, whose agent model is upgraded repeatedly and is increasingly a closed-source frontier model served through an API that cannot be fine-tuned. A weight-tuned policy would need re-training on every upgrade, if it could be trained at all, whereas our natural-language heuristics carry across model generations intact.


\revise{We evaluate EnumGRPO on four SWAN databases~\cite{zhao2025hybrid}. On SWAN, EnumGRPO achieves 35.4\% execution accuracy, outperforming Agentic Text2SQL, Agentic BlendSQL, and oracle-assisted LOTUS and Palimpzest. Its LLM-operator cost is only \$0.011 per query, 317$\times$ lower than Agentic BlendSQL and 55$\times$ lower than Palimpzest. Without retraining, our system transfers the SWAN-learned experiences to a new LLM family, to SemBench's text-only Movie workload~\cite{lao2025sembench}, and to the purely relational Spider~\cite{yu2018spider} and BIRD~\cite{li2024bird} workloads. On SemBench Movie, it reaches 0.713 Quality, matching oracle-program BlendSQL at 23$\times$ lower LLM-operator cost and reducing cost by 134$\times$ and 210$\times$ relative to LOTUS and Palimpzest. On Spider and BIRD, it shortens trajectories without sacrificing accuracy or invoking unnecessary semantic operators.}

In summary, this paper makes the following contributions:
\begin{itemize}[leftmargin=*]
    \item We formulate \textbf{agentic query execution} and its query optimization as an \textbf{agent workflow optimization} problem, in which agent-based planning is interleaved with execution over intermediate results. It generalizes classical query optimization from cost-only search to joint quality-cost search over dynamically constructed plans. 
    \item We characterize the plan space of agentic query execution along five axes: execution paradigm, operator type, operator placement, selectivity scope, and projection width, analogous to classical query optimization.
    \item We propose \textbf{EnumGRPO}, a learned optimizer that uses structured plan enumeration during a learning stage paired with in-context reinforcement learning to distill reusable query optimization heuristics, enabling better workflow construction at query time without model retraining.
    \item \revise{System comparisons, ablations, and transfer experiments show that agentic query execution with EnumGRPO improves the quality--cost frontier on SWAN, transfers across LLM backbones and to a new semantic workload on SemBench, and avoids LLM operators on Spider and BIRD when SQL suffices.}
\end{itemize}

\section{Problem Formulation}
\label{sec:formulation}

We first formalize \name as a paradigm in which agent-based planning is interleaved with execution over an extended relational algebra: after each step, the system observes new intermediate results and decides which operator to apply next, whether to stay inside the DBMS, and when to pay for approximate LLM-backed semantic processing.

\subsection{Task Setting}
\label{subsec:task-setting}
Let $\mathcal{D}$ denote a relational database instance with base relations $\{R_i\}_{i=1}^m$ and schemas $\{\mathsf{sch}(R_i)\}$. An analytics task is specified by a natural-language query $q$, optionally with additional textual context such as column descriptions or domain-specific conventions. The system is equipped with a toolset:
\[
K = K_{\textsc{sql}} \cup K_{\textsc{llm}} \cup K_{\textsc{aux}},
\]
where $K_{\textsc{sql}}$ contains DBMS-resident relational processing, $K_{\textsc{llm}}$ contains LLM-backed semantic operators, and $K_{\textsc{aux}}$ contains supporting primitives such as schema inspection, sampling, or summarization.

\revise{Execution proceeds until the system terminates and produces a final output $y$.} At turn $t$, the system maintains a \emph{workspace} $W^t$ containing named intermediate relations, schema and catalog fragments, lightweight statistics or samples, and the execution trace accumulated so far.

\subsection{Logical Operators, Physical Tools, and Actions}
\label{subsec:operators-actions}
\paragraph{Logical operators.}
Mirroring the logical-to-physical separation in classical query processing~\cite{graefe1995cascades}, we distinguish logical operators from their physical tool implementations. A logical operator is a typed transformation over workspace artifacts. This includes relational operators expressible in SQL, such as selection, projection, join, grouping, and aggregation, as well as semantic operators whose behavior is realized through LLM inference~\cite{DBLP:journals/corr/abs-2407-11418, lao2025sembench, patel2025semanticoptimization}.

Let $\mathcal{R}$ denote the set of all possible finite relations, let $\mathcal{R}_{\textsc{score}}$ denote relations augmented with a real-valued score attribute, and let $\mathbf{1}$ denote singleton outputs. Representative semantic operators include:
\begin{align*}
\texttt{llm\_map} &:\ \mathcal{R} \rightarrow \mathcal{R}  \ \ \text{(transformation)}\\
\texttt{llm\_filter} &:\ \mathcal{R} \rightarrow \mathcal{R} \ \ \text{(selection)}\\
\texttt{llm\_rank} &:\ \mathcal{R} \rightarrow \mathcal{R}_{\textsc{score}}  \\
\texttt{llm\_reduce} &:\ \mathcal{R} \rightarrow \mathbf{1} \\
\texttt{llm\_join} &:\ \mathcal{R} \times \mathcal{R} \rightarrow \mathcal{R}.
\end{align*}
Unlike their relational counterparts, semantic operators are \emph{approximate}: the quality of their output depends on the cardinality and content of the input relation, so placing the same logical operator at different points in a workflow can yield different-quality results.

\paragraph{Physical tools.}
A tool $k \in K$ is a callable physical implementation of one or more logical operators. A SQL tool may compile directly to DBMS execution, while an LLM tool fixes a model family, serving configuration, prompt template, batching policy, and any surrounding SQL pre- or post-processing. A tool may implement a single operator or a macro that expands into a short operator pipeline.

\paragraph{Actions.}
An action is one concrete invocation
\[
a_t = (k_t, \eta_t, w_t),
\]
where $k_t \in K$ is the chosen tool, $\eta_t$ are its parameters, and $w_t \subseteq W^t$ identifies the workspace artifacts consumed by the action. Executing $a_t$ produces updated artifacts and yields a new workspace $W^{t+1}$. SQL actions are deterministic for a fixed database snapshot, while LLM-tool actions can be stochastic and approximate.

\subsection{Agent Workflow Optimization}
\label{subsec:workflow-optimization}
Within this new execution paradigm, query optimization becomes \emph{agent workflow optimization}. Unlike classical optimization over equivalent plans, the optimizer must now navigate plans that differ in both cost and answer quality.

\paragraph{State and feasible actions.}
At query time, the optimizer observes only the current task and workspace, without any reference answer. We define the state at turn $t$ as
\[
s_t = (q, W^t).
\]
Let $\mathcal{S}$ denote the set of reachable states. Given $s_t \in \mathcal{S}$, the optimizer selects from a feasible action set $\mathcal{A}(s_t)$, containing all tool invocations whose required inputs are present in $W^t$, together with a distinguished action \textsc{stop} that terminates execution and returns output $y$ from the current workspace.

\paragraph{Workflow cost.}
Each non-terminal action incurs a nonnegative cost $c(s_t,a_t)$, which may reflect latency, computation overhead, or monetary expense, depending on the deployment setting. Since a tool can expand into multiple physical substeps, we interpret
\[
c(s_t,a_t) \;=\; c_{\textsc{sql}}(s_t,a_t) + c_{\textsc{llm}}(s_t,a_t) + c_{\textsc{aux}}(s_t,a_t),
\]
where the three terms capture DBMS work, LLM usage, and auxiliary processing, respectively. In practice, $c_{\textsc{llm}}$ dominates the other two by orders of magnitude.

\paragraph{Workflow utility.}
Let $U(y; q, \mathcal{D}) \in \mathbb{R}$ denote the utility of the final output $y$, measuring answer quality for the user task. We leave $U$ abstract: it may correspond to exact-match correctness, a soft similarity score, or any task-specific quality measure. Without a reference answer, $U$ is unobservable at query time, so the optimizer must trade estimated quality against resource consumption.

\paragraph{Policy and objective.}
A policy $\pi(a \mid s)$ defines a distribution over feasible next actions given the current state. Executing $\pi$ induces an adaptive workflow
\[
\tau = (a_1, a_2, \ldots, a_T, \textsc{stop}),
\]
\revise{where $T$ denotes the number of non-terminal actions executed before termination.} Later actions depend on intermediate relations materialized during earlier turns. Each workflow produces a trajectory reward that trades output quality against cumulative cost:
\begin{equation}
\label{eq:reward}
G(\tau) \;=\; U(y;\, q, \mathcal{D}) \;-\; \lambda \sum_{t=1}^{T} c(s_t,a_t),
\end{equation}
for a tradeoff parameter $\lambda \ge 0$. The primary optimization problem is to maximize the expected reward under policy $\pi$:
\begin{equation}
\label{eq:objective}
\max_{\pi}\;\; J(\pi)
\;=\;
\max_{\pi}\;\;
\mathbb{E}_{\tau \sim \pi}\!\left[G(\tau)\right].
\end{equation}
Concretely, the policy must make structural planning decisions at each step, such as which operator to invoke and where to place it, while accounting for the fact that LLM-backed operators are both expensive and approximate.

\begin{table*}[t]
  \centering
  \caption{The plan enumeration space for \name. \revise{\textsc{n.a.} denotes an axis that does not apply to code-driven plans.}}
  \label{tab:plan-axes}
  \resizebox{\textwidth}{!}{%
  \begin{tabular}{@{}llll@{}}
  \toprule
  \textbf{Axis} & \textbf{Values} & \textbf{Description} & \textbf{Optimizer analogue} \\
  \midrule
  Execution paradigm
    & \textsc{data-driven} / \textsc{code-driven}
    & \revise{Choose direct LLM inference or executable-program synthesis.}
    & JIT query compilation \cite{neumann2011efficiently} \\
  Operator type
    & \revise{\textsc{scalar} / \textsc{aggregate} / \textsc{n.a.}}
    & \revise{For data-driven plans, choose row-wise or set-wise LLM inference.}
    & Join algorithm selection \cite{chaudhuri1998overview} \\
  Operator placement
    & \textsc{pre-} / \textsc{post-aggregation}
    & \revise{Place semantic processing before or after SQL aggregation.}
    & Predicate pushdown \cite{selinger1979} \\
  Selectivity scope
    & \textsc{full} / \textsc{targeted}
    & \revise{Apply semantic processing to the full or a targeted working set.}
    & Access-path \& materialization \cite{selinger1979,menon2017relaxed} \\
  Projection width
    & \revise{\textsc{narrow} / \textsc{wide} / \textsc{n.a.}}
    & \revise{For data-driven plans, control the row context exposed to the LLM.}
    & Late materialization \cite{abadi2008column} \\
  \bottomrule
  \end{tabular}%
  }
\end{table*}

\section{Method}
\label{sec:method}

\subsection{Plan Space of Agentic Query Execution}
\label{subsec:plan-space}

\revise{Traditional query optimization improves execution plans through selection of physical implementations and logical plan transformations~\cite{ioannidis1996query,chaudhuri1998overview}. Physical implementation choices determine how operations are executed, while transformations such as operator reordering and selection and projection pushdown change the intermediate data processed at each stage~\cite{jarke1984query,hellerstein1993predicate}. Viewed together, these choices suggest a common optimization principle: jointly consider how a computation is implemented, where it appears in the plan, and what data it consumes.}


\revise{We apply this principle to semantic processing through five planning axes that capture the execution choices considered in this work. Execution paradigm and operator type characterize \textbf{how} semantic computation is realized, distinguishing execution mechanism from invocation granularity. Operator placement characterizes \textbf{where} semantic processing occurs relative to SQL aggregation. Selectivity scope and projection width characterize \textbf{what data} it consumes, controlling respectively which rows are processed and which attributes are exposed as context. During learning, each assignment of axis values guides the construction of a candidate workflow (Section~\ref{subsec:plan-enum}); Table~\ref{tab:plan-axes} summarizes their conditional applicability.

\revise{These choices affect both LLM cost and answer quality. For a fixed model, we use input-token volume as a first-order proxy for input-side cost of a single execution step:
\[
L_{\mathrm{in}} \approx N_{\text{calls}} \cdot t_{\text{base}} + N_{\text{rows}} \cdot t_{\text{row}},
\]
where $N_{\text{calls}}$ is the number of LLM-operator invocations, $N_{\text{rows}}$ counts row occurrences across all invocations, and $t_{\text{base}}$ and $t_{\text{row}}$ are the average non-row and per-row input tokens, respectively. Execution paradigm affects both $N_{\text{calls}}$ and $N_{\text{rows}}$; operator type primarily affects $N_{\text{calls}}$; operator placement and selectivity scope primarily affect $N_{\text{rows}}$; and projection width affects $t_{\text{row}}$. For example, code-driven execution can replace repeated inference with program synthesis, reducing token volume but potentially sacrificing quality when the semantic rule is difficult to capture.}}

\begin{example}
\label{ex:running}
A table \texttt{Companies} stores company records but has no \texttt{sector} column:

\smallskip
\centering
\begin{small}
\begin{tabular}{clrc}
\toprule
\textbf{id} & \textbf{name} & \textbf{revenue} & \textbf{headquarters} \\
\midrule
1 & Medtronic  & 31{,}200M & Minneapolis, MN \\
2 & Genesis    & 8{,}400M  & Dearborn, MI \\
3 & Stryker    & 19{,}800M & Kalamazoo, MI \\
4 & Apex Solar & 2{,}100M  & Austin, TX \\
5 & \ldots & \ldots & \ldots \\
\bottomrule
\end{tabular}
\end{small}

\smallskip
\raggedright
A user asks: \emph{``What is the total revenue of healthcare companies?''} Answering this query requires the system to infer each company's sector from its name and other attributes using world knowledge not present in the database.
\end{example}

\paragraph{Execution paradigm.}
\revise{Semantic computation presents a compile-vs.-interpret choice analogous to compiled query execution~\cite{neumann2011efficiently}. It can be performed either through direct LLM inference over data~\cite{patel2025semanticoptimization} or through an LLM-synthesized executable program~\cite{sqlmorpher,pal}. In the \emph{data-driven} paradigm, an LLM operator reads relation contents and directly produces inferred values, selections, or answers. In the \emph{code-driven} paradigm, the planner uses the schema and a small sample to synthesize a task-specific deterministic SQL program---for example, a \texttt{CASE WHEN} mapping, regular-expression transformation, or SQL UDF---which the DBMS executes over the working relation. This shifts bulk processing to a deterministic runtime and avoids repeated LLM inference over individual records. Thus, execution paradigm determines whether semantic computation is performed through direct inference or synthesized code.}
\begin{mdframed}[style=exbar]
\textsf{\textbf{In Example~\ref{ex:running}:}}
\begin{description}[nosep,leftmargin=1em,font=\normalfont\scshape]
  \item[data-driven] \revise{The LLM directly produces sector labels from company tuples.}
  \item[code-driven] The LLM synthesizes a \texttt{CASE WHEN} mapping company names to sectors and executes it as pure SQL.
\end{description}
\end{mdframed}

\paragraph{Operator type.}
A query optimizer chooses among join algorithms with different cost profiles: a nested-loop join probes its inner input for each outer tuple, whereas a hash join builds a hash table over one input and probes it with the other \cite{chaudhuri1998overview}. \revise{Analogously, given the rows selected for semantic processing, an LLM operator can process them independently or jointly}. A \emph{scalar} operator, \eg \texttt{llm\_map}, applies the LLM to each row independently, producing a per-row result (an inferred attribute, a boolean label, a normalized value). An \emph{aggregate} operator, \eg \texttt{llm\_reduce}, passes an entire relation or a grouped partition to the LLM in a single call, producing an aggregate or summary. Scalar processing offers fine-grained control and is robust when rows require different reasoning, while aggregate processing amortizes prompt overhead and is more effective when the LLM can reason about the full context simultaneously. \revise{Thus, invocation granularity trades LLM-call cost against the reasoning benefits of independent or joint processing.}
\begin{mdframed}[style=exbar]
\begin{description}[nosep,leftmargin=1em,font=\normalfont\scshape]
  \item[scalar] Classifies each company's sector one row at a time.
  \item[aggregate] Receives all company names at once and returns only those in healthcare, cross-referencing entries to recognize that ``Medtronic'' and ``Stryker'' both belong to the same sector.
\end{description}
\end{mdframed}

\paragraph{Operator placement.}
Predicate pushdown is a cornerstone of relational optimization: the optimizer pushes selective predicates as early as possible in the plan tree, but it cannot push a predicate below the operator that produces the column it references \cite{selinger1979}. \revise{The same dependency principle applies to semantic processing. \emph{Pre-aggregation} placement is required when downstream SQL depends on a semantically derived value, whereas \emph{Post-aggregation} placement can defer semantic processing until SQL has reduced the working set. Thus, operator placement trades earlier access to semantic values against lower semantic-processing cost.}
\begin{mdframed}[style=exbar]
\begin{description}[nosep,leftmargin=1em,font=\normalfont\scshape]
  \item[pre-aggregation] \revise{
  The healthcare filter depends on the inferred \texttt{sector}, so semantic processing must classify companies before \texttt{SUM(revenue)} can be computed.}
  \item[post-aggregation] \revise{If a variant query asked whether the portfolio's total revenue is exceptionally large for its market, SQL could first compute \texttt{SUM(revenue)}, after which the LLM interprets that aggregate using external domain knowledge.}
\end{description}
\end{mdframed}

\paragraph{Selectivity scope.}
Access-path selection determines how much of a relation the engine actually reads: an index-restricted scan touches only qualifying rows, while a full scan reads the entire table \cite{selinger1979}. Materialization boundaries further control which intermediate results are physically produced and which are pipelined \cite{menon2017relaxed}. \revise{In agentic workflows, the analogous decision is how much of the working set undergoes semantic processing. Under \emph{full} scope, the entire filtered relation is processed. Under \emph{targeted} scope, the planner uses a tight SQL predicate to restrict processing to the minimum set of rows required to answer the query.}
\begin{mdframed}[style=exbar]
\begin{description}[nosep,leftmargin=1em,font=\normalfont\scshape]
  \item[full] Every company is sent to the LLM for sector classification.
  \item[targeted] A SQL name predicate resolves easy cases without the LLM. Only the remaining ambiguous rows are classified.
\end{description}
\end{mdframed}

\paragraph{Projection width.}
Column stores defer tuple reconstruction until after all predicates have been evaluated, avoiding the cost of materializing columns that are never needed \cite{abadi2008column}. This late-materialization principle extends directly to LLM operator inputs. Under \emph{narrow} projection, only the target column and the primary key are included in each LLM call, minimizing per-call token cost. Under \emph{wide} projection, the full row or a semantically relevant subset of columns is included, giving the LLM the surrounding context it may need to make accurate inferences.
\begin{mdframed}[style=exbar]
\begin{description}[nosep,leftmargin=1em,font=\normalfont\scshape]
  \item[narrow] Only \texttt{name} and the primary key are sent. Token cost is minimal.
  \item[wide] Including \texttt{headquarters} lets the LLM disambiguate: ``Genesis'' in ``Dearborn, MI'' is likely automotive, not healthcare.
\end{description}
\end{mdframed}

\paragraph{\revise{Adaptive meta-plan.}}
\revise{The five axes yield 20 fixed strategies. Under data-driven execution, the four remaining binary axes all apply, giving $2^4=16$ combinations. Under code-driven execution, operator type and projection width do not apply, leaving $2^2=4$ combinations over placement and selectivity scope. We add one adaptive strategy that leaves the axis values unspecified and lets the agent choose them at runtime after inspecting intermediate data. Thus, the plan space contains 21 assignments.} This mirrors the Eddies principle \cite{avnur2000eddies} of deferring routing decisions to runtime based on observed selectivities. The learned heuristics in $\varepsilon$ use observed data characteristics to rank axis values.

\paragraph{Contrasting workflows.}
The compositional power of the plan space becomes visible when different axis combinations are applied to Example~\ref{ex:running}. Consider three workflows:
\begin{enumerate}
\item \textbf{Conservative:} \textsc{data-driven}, \textsc{scalar}, \textsc{pre-aggregation}, \textsc{full} scope, \textsc{wide} projection. The agent infers the sector of every company row-by-row, including \texttt{headquarters} for context, then runs \texttt{GROUP BY sector} and \texttt{SUM(revenue)}. This yields the highest accuracy but incurs $O(|R|)$ LLM calls with wide prompts, making it the most expensive option.
\item \textbf{Aggressive:} \revise{\textsc{code-driven}, \textsc{pre-aggregation}, \textsc{full} scope. The agent uses a small planning sample to synthesize a \texttt{CASE WHEN} rule, then applies that rule natively to the full working relation.} Total cost is $O(1)$ LLM calls, but accuracy degrades for companies the rule cannot cover.
\item \textbf{Balanced:} \textsc{data-driven}, \textsc{aggregate}, \textsc{pre-aggregation}, \textsc{targeted} scope, \textsc{narrow} projection. \revise{The agent first uses SQL predicates to narrow the candidate companies,
then sends only their names to the LLM in one batch to identify which belong to healthcare. It then filters to those companies and computes \texttt{SUM(revenue)}. Compared with the conservative plan, this reduces both the number of rows and the amount of context sent to the LLM, while still allowing the LLM to reason over multiple candidates jointly.}
\end{enumerate}
All three plans are valid for the same query. \revise{Which workflow offers the best quality--cost tradeoff} depends on the data distribution, table size, and the user's quality-cost preference. \revise{Each axis assignment provides a planning strategy that the agent adapts to the current task by choosing concrete tools, SQL expressions, predicates, arguments, and intermediate relations. The assignment therefore guides workflow construction while preserving task-specific planning.}

\begin{figure*}[t]
  \centering
  \includegraphics[width=\textwidth]{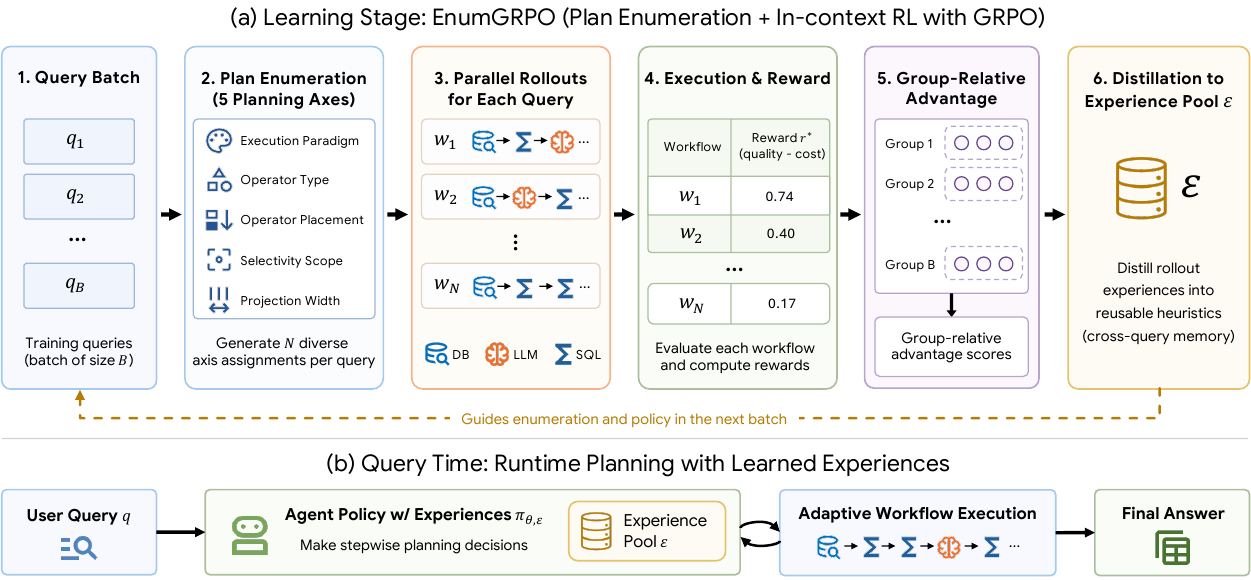}
  \caption{Overview of EnumGRPO. \textbf{(a)}~During the learning stage, queries are processed in batches: a PlanEnumerator generates $N$ diverse axis assignments per query, each seeded workflow is executed in parallel and scored by a composite reward to compute group-relative advantages, then distilled into reusable heuristics stored in the experience pool $\varepsilon$. \textbf{(b)}~At query time, the agent applies the learned $\varepsilon$ to make stepwise planning decisions and executes an adaptive workflow without enumeration.}
  \label{fig:enumgrpo}
\end{figure*}

\paragraph{\revise{From plan space to learned optimization.}}
\revise{Unlike classical query optimization, where alternative plans are expected to preserve query results, different axis combinations in our setting can vary in both cost and answer quality. Because these effects depend on observed data, a fixed cost model alone cannot reliably rank their quality--cost tradeoffs. We therefore learn from executed workflows which planning choices perform well under different conditions, following the learned-optimization direction exemplified by Bao~\cite{bao2021}. }

\subsection{\revise{Learned Optimization via In-context RL}}
\label{subsec:icrl-formulation}
\revise{Following this learned-optimizer perspective, we formulate the learning problem for the objective defined in Section~\ref{sec:formulation}.} The optimization problem is inherently sequential: each planning decision changes the workspace and thus the set of useful downstream actions. Improving the planner therefore requires credit assignment over entire workflow trajectories rather than supervision on a single next action, which suggests a reinforcement learning perspective. \revise{The workflow forms an episodic Markov Decision Process (MDP):}
\[
\revise{\mathcal{M} = (\mathcal{S}, \mathcal{A}, P, \rho_0),}
\]
\revise{where $P(s_{t+1} \mid s_t, a_t)$ is the transition kernel induced by executing $a_t$ and updating the workspace, and $\rho_0$ is the initial state determined by the database instance and the task query.}

During the offline learning stage, we have access to tasks with reference outputs $y^*$, which lets us instantiate the abstract utility $U$ with a concrete benchmark-specific score $S(y, y^*;\, q, \mathcal{D})$ that compares the predicted output to the reference. The trajectory reward from Eq.~\eqref{eq:reward} becomes
\[
G(\tau) \;=\; S(y, y^*;\, q, \mathcal{D}) \;-\; \lambda \sum_{t=1}^{T} c(s_t,a_t),
\]
where the tradeoff parameter $\lambda$ may differ between the learning stage and query time. Maximizing $J(\pi)$ over this reward would normally require updating model parameters.

However, updating an LLM's weights through gradient-based training is expensive, and is infeasible when the model is accessed only through an API. We therefore adopt \emph{in-context reinforcement learning} (ICRL): the model parameters $\theta$ remain fixed, and policy improvement operates through the prompt context supplied to the model \cite{dherin2025implicit,song2026reward,DBLP:journals/corr/abs-2507-19457}.

\revise{Let $\mathcal{C}$ denote the ground set of admissible context elements, such as planning guidelines, demonstrations, or instructions. We represent the learnable portion of the prompt context as an experience pool $\varepsilon \subseteq \mathcal{C}$; the fixed portion of the system prompt is omitted from the notation. Prepending $\varepsilon$ to the system prompt induces the policy $\pi_{\theta,\varepsilon}(a_t \mid s_t)$. The ICRL objective therefore restricts optimization to the space of experience pools:}
\begin{equation}
\label{eq:icrl-objective}
\revise{\max_{\varepsilon \,\subseteq\, \mathcal{C}}\;\; J(\pi_{\theta,\varepsilon})}
\;=\;
\revise{\max_{\varepsilon \,\subseteq\, \mathcal{C}}\;\;
\mathbb{E}_{\tau \sim \pi_{\theta,\varepsilon}}\!\left[G(\tau)\right].}
\end{equation}
This is policy optimization over a restricted search space, avoiding the substantial costs of gradient-based parameter updates. Equation~\eqref{eq:icrl-objective} is solved offline, where oracle outputs $y^*$ provide the concrete score $S$ needed to evaluate trajectories. \revise{At query time, the system applies the learned experience pool $\varepsilon$ to construct workflows without access to ground-truth answers.}

\subsection{Learning to Navigate the Plan Space}
\label{subsec:enumgrpo-overview}

\revise{With the plan space and ICRL objective defined, we now introduce the learning mechanism. It must (i)~generate diverse candidate behaviors for the same query, (ii)~score them, and (iii)~distill the relative outcomes into reusable guidance.}

\paragraph{GRPO.}
Group Relative Policy Optimization (GRPO)~\cite{shao2024deepseekmath} provides exactly this structure. For each input, GRPO samples a group of $N$ rollouts, scores them, and normalizes rewards within the group so that above-average attempts receive positive advantage and below-average ones receive negative advantage. This group-relative signal fits the \name setting particularly well. A healthcare-revenue query over a 10-row table and one that classifies sectors across 10{,}000 rows \revise{can exhibit substantially different reward distributions}. Within-group normalization absorbs this variance and yields a stable learning signal without global reward calibration. More importantly, since each rollout group targets one query, its contrastive comparison isolates plan-axis effects and aligns with the credit-assignment problem in Section~\ref{subsec:plan-space}: if a \textsc{code-driven}/\textsc{pre-agg} workflow scores higher than a \textsc{data-driven}/\textsc{post-agg} alternative on the same data, the advantage \revise{reflects the relative performance of the structural alternatives under the same query and data}.

\paragraph{From GRPO to EnumGRPO.}
Standard GRPO relies on temperature sampling to produce diverse rollouts. \textbf{EnumGRPO} replaces this undirected diversity with structured plan enumeration, seeding each rollout in a group with a distinct axis-value assignment from the plan space. This guarantees that rollout groups cover meaningfully different structural alternatives, producing stronger contrastive signal for credit assignment. \revise{EnumGRPO realizes Eq.~\eqref{eq:icrl-objective} without parametric updates: the LLM planner's weights remain fixed, while policy improvement operates entirely through group-relative updates to the evolving experience pool $\varepsilon$.}

The algorithm (Figure \ref{fig:enumgrpo}) has three major components. First, a \emph{plan enumerator} exploits the structured plan space from Section~\ref{subsec:plan-space} to generate \revise{diverse strategy assignments} per query, guaranteeing that rollout groups cover different regions of the axis space rather than collapsing under temperature noise alone. Second, a \emph{multi-objective reward} scores each workflow on both answer quality and LLM-operator cost, instantiating the trajectory reward $G(\tau)$ from the formulation with quality signals at multiple granularities to avoid reward sparsity. Third, a \emph{four-stage experience distillation pipeline} converts group-relative advantage signals into reusable planning heuristics that update $\varepsilon$ across learning batches.

\subsubsection{Plan Enumeration}
\label{subsec:plan-enum}

To seed each rollout group with diverse strategies, the \texttt{PlanEnumerator} selects $N$ axis-value assignments before execution begins. Given the query, the database schema, and a small row sample, a single LLM call proposes ranked assignments as structured JSON. \revise{We discard duplicate structural tuples and require coverage of at least one scalar-map plan, one aggregate-reduce plan, and one code-driven plan. If a category is absent from the first $N$ unique assignments, its highest-ranked remaining candidate replaces the lowest-ranked selected assignment. Infeasible code-driven operator-type and projection-width fields are omitted.} Each assignment is then injected into the agent prompt as a soft recommendation that the agent may override when the data warrants a different approach.

\subsubsection{Multi-Objective Reward}
\label{subsec:reward}

Each rollout receives a composite reward combining answer quality at three granularities with a cost penalty. A binary exact-match score alone is too sparse: for harder queries, most rollouts score zero, collapsing advantages and making near-correct workflows indistinguishable from completely wrong ones. We therefore decompose quality into result-level, tuple-level, and cell-level components that credit partially correct outputs. 

\paragraph{Execution Accuracy (EX).}
Let $\hat{R}$ denote the agent's output relation and $R^*$ the ground-truth relation. The coarsest signal is exact-match correctness~\cite{DBLP:journals/pvldb/LiLCLT24,DBLP:journals/pvldb/KimSHL20,DBLP:journals/pvldb/GaoWLSQDZ24}:
\begin{equation}
\label{eq:ex}
\text{EX}(\hat{R}) = \mathbf{1}[\hat{R} = R^*].
\end{equation}
EX provides an unambiguous task-completion signal but, used alone, would yield a reward of zero for any rollout that retrieves most rows correctly yet misses or misformats a single value.

\paragraph{Tuple-level F1 (TupleF1).}
To reward partial overlap at the record level, we use a tuple-level F1 score drawn from entity resolution evaluation~\cite{DBLP:journals/pvldb/MenestrinaWG10,DBLP:journals/pvldb/KopckeTR10,DBLP:journals/pvldb/0001LSDT20}:
\begin{equation}
\label{eq:tuplef1}
\text{TupleF1}(\hat{R}) = \frac{2|\hat{R} \cap R^*|}{|\hat{R}| + |R^*|},
\end{equation}
where $\hat{R} \cap R^*$ counts tuples matched by primary key and all attribute values. TupleF1 gives credit when the agent retrieves a correct subset or superset of tuples, producing a nonzero reward even when the result set is not an exact match.

\paragraph{Cell-level F1 (CellF1).}
The finest-grained signal operates at the individual cell level~\cite{DBLP:journals/pvldb/TataPWCNG21,DBLP:journals/corr/abs-2407-11418,DBLP:journals/pvldb/EltabakhNAOT24}:
\begin{equation}
\label{eq:cellf1}
\text{CellF1}(\hat{R}) = \frac{2\sum_{c}|\hat{R}_c \cap R^*_c|}{\sum_{c}(|\hat{R}_c| + |R^*_c|)},
\end{equation}
where $\hat{R}_c$ and $R^*_c$ are the multisets of values in column $c$. CellF1 surfaces differences between rollouts that are invisible at the tuple level, for instance when two workflows return the same rows but one infers a cell value correctly while the other does not.

\paragraph{Composite reward.}
The three granularities are combined into a single quality score for each rollout producing output $\hat{R}$ to query $q$:
\[
r = w_{\textsc{ex}}\cdot\text{EX}(\hat{R}) + w_{\textsc{tuple}}\cdot\text{TupleF1}(\hat{R}) + w_{\textsc{cell}}\cdot\text{CellF1}(\hat{R})
\]

Since query complexity changes token scale, we normalize cost within each rollout group. For rollout $i$, let $t_i$ and $s_i$ denote its token and tool-call counts:
\[
\rho_i = w_{\textsc{token}} \cdot \frac{t_i - \min_j t_j}{\max_j t_j - \min_j t_j} + w_{\textsc{step}} \cdot \frac{s_i - \min_j s_j}{\max_j s_j - \min_j s_j}.
\]
A zero-range normalization term is defined as zero. The final reward is $r_i^* = r_i - w_{\textsc{cost}} \cdot \rho_i$, with the small $w_{\textsc{cost}}$ making cost a tiebreaker among similarly accurate rollouts.

\subsubsection{Group-Relative Advantage}
\label{subsec:advantage}

Following GRPO~\cite{shao2024deepseekmath, cai2025trainingfree}, advantages are computed group-relatively for each query. For $N$ rollouts with rewards $r_1^*, \ldots, r_N^*$:
\[
A_i = \frac{r_i^* - \mu}{\max(\sigma,\epsilon)}, \quad \mu = \frac{1}{N}\sum_j r_j^*, \quad \sigma = \sqrt{\frac{1}{N}\sum_j (r_j^* - \mu)^2},
\]
where $\epsilon>0$ is a small numerical-stability constant.

Alongside $A_i$, we compute $A_i^{\textsc{util}}$ and $A_i^{\textsc{cost}}$ by $z$-normalizing quality scores and negated cost ranks, respectively, to support plan-axis attribution during distillation.

\begin{algorithm}[t]
\caption{EnumGRPO Learning Loop}
\label{alg:enumgrpo}
\begin{algorithmic}[1]
\Require database  $\mathcal{D}$, query set $\mathcal{Q}$, enumeration axes $\mathcal{L}$
\State $\varepsilon \leftarrow \emptyset$ \Comment{\textcolor{gray}{\textit{init experience pool}}}
\For{each batch $\{q_1, \ldots, q_B\} \subseteq \mathcal{Q}$}
  \For{each query $q_i$}
    \State $\{\phi_j\}_{j=1}^N \!\leftarrow\! \texttt{PlanEnumeration}(q_i, \mathcal{D}, \mathcal{L}, N)$
    \Statex \hspace{\algorithmicindent}\hspace{\algorithmicindent}\Comment{\textcolor{gray}{\textit{enumerate $N$ diverse axis assignments}}}
    \For{$j = 1$ to $N$ \textbf{in parallel}}
      \State $\tau_{ij} \leftarrow \texttt{Rollout}(q_i,\, \phi_j,\, \varepsilon)$ \Comment{\textcolor{gray}{\textit{execute with hint $\phi_j$}}}
      \State $r_{ij} \leftarrow \texttt{Reward}(\tau_{ij})$ \Comment{\textcolor{gray}{\textit{multi-objective score}}}
    \EndFor
    \State Compute group-relative advantage $A_{ij}$ from $r_{i1},\ldots,r_{iN}$
  \EndFor
  \State $\varepsilon \leftarrow \texttt{Distill}(\{(\tau_{ij}, A_{ij})\},\, \varepsilon)$
  \Comment{\textcolor{gray}{\textit{experience distillation}}}
\EndFor
\State \Return experience pool $\varepsilon$
\end{algorithmic}
\end{algorithm}

\subsubsection{Experience Distillation}
\label{subsec:distill}

The core policy-improvement mechanism is a four-stage LLM pipeline that converts group-relative advantages into reusable planning heuristics. No gradient steps are taken: the ``update'' to the policy is a textual edit to the persistent experience pool $\varepsilon$ that conditions all subsequent rollouts. The four stages map rollout observations to a global pool update:

\paragraph{Stage 1: Trajectory summarization.}
For each rollout, an LLM compresses the execution trajectory into a compact behavioral description. The summary receives the query, output, decomposed advantages ($A_i$, $A_i^{\textsc{util}}$, $A_i^{\textsc{cost}}$), and a per-step execution profile that records tool arguments, row cardinality, and LLM token breakdown. This profile supplies a dense, step-level signal that complements the end-to-end reward: it makes localized inefficiencies visible, such as a large \texttt{llm\_map} input that could have been pre-filtered by SQL, even when the final answer happens to be correct.

\paragraph{Stage 2: Contrastive analysis.}
For each query, all rollout summaries are presented together, annotated with their advantages and plan-axis labels. The LLM contrasts high-advantage and low-advantage attempts and proposes generalizable experiences: short, action-oriented guidelines such as ``apply SQL narrowing before \texttt{llm\_map} when the inferred column is not on the join key.'' The axis labels allow the LLM to reason directly about which structural choice drove the outcome.

\paragraph{Stage 3: Pool update.}
Proposed experiences are compared against the current pool. An LLM assigns one of three labels: ADD for a genuinely new proposal, UPDATE for a refinement of an existing entry, and DELETE for a contradiction.

\paragraph{Stage 4: Consolidation.}
One consolidation call merges all per-query operations, deduplicates semantically overlapping additions, and ensures global consistency. The resulting pool is written to disk and appended to the agent's system prompt as the updated $\varepsilon$ that parameterizes the policy $\pi_{\theta,\varepsilon}$ in subsequent rollouts.
\subsection{Learning and Execution}
\label{subsec:training-loop}

Algorithm~\ref{alg:enumgrpo} gives the full EnumGRPO learning loop. Queries are processed in batches of $B$. For each batch, $N$ rollouts are run per query in parallel, each seeded with a different plan-axis assignment from the enumerator. Rewards are computed, advantages are derived, and the four-stage pipeline updates the experience pool. The updated pool is injected into the agent prompt for the next batch.

At query time, the learned experience pool $\varepsilon$ is prepended to the agent's system prompt. No plan enumeration is performed: the agent receives a new query and its database, plans and executes a workflow step by step using the heuristics in $\varepsilon$, and returns a final answer. The learned experiences thus serve as a lightweight planning model that guides the agent's axis-level decisions without any runtime overhead beyond the additional prompt tokens.

\section{Experimental Setup}
\label{sec:exp-setup}


\subsection{Evaluation}
\label{subsec:evaluation}

\paragraph{Benchmark.} \revise{Our primary evaluation uses} a benchmark whose queries cannot be answered by relational
operators alone but instead demand semantic reasoning. 
We evaluate on SWAN~\cite{zhao2025hybrid}, a cross-domain benchmark of 120 beyond-database questions over four real-world SQLite databases derived from BIRD~\cite{li2024bird}. Each question has partial information grounded in the relational tables but additionally requires world knowledge that is absent from the schema (e.g., mapping a superhero name to its publisher, or a zip code to a city). SWAN creates this setting by removing selected columns and join paths from the complete BIRD databases. The system sees only the incomplete schema, while ground-truth answers are defined by executing gold SQL on the original databases.
We additionally drop columns whose values are trivially 
recoverable via a single join, ensuring no purely 
relational path to the answer. Table~\ref{tab:swan-stats} summarizes the database statistics. We randomly split the 120 questions into 40 learning and 80 evaluation queries.

\revise{For cross-workload transfer, we first include the \emph{Movie} scenario from SemBench~\cite{lao2025sembench} as a separate semantic query workload. Its ten queries over 2{,}000 reviews all
require semantic processing. The remaining scenarios in SemBench require multimodal processing that is outside our
system's scope. For pure-SQL workloads, we use 166
\emph{extra-hard} Spider~\cite{yu2018spider} queries across 16 databases and 231
\emph{challenging} BIRD~\cite{li2024bird} queries across 11 databases. For these two benchmarks, we convert the SQLite databases and reference queries to
DuckDB, then audit execution equivalence with the originals.}

\begin{table}[t]
  \centering
  \caption{Statistics of databases in SWAN~\cite{zhao2025hybrid}.}
  \label{tab:swan-stats}
  \begin{tabular}{lrrr}
  \toprule
  \textbf{Database} & \textbf{Tables} & \textbf{Rows/Table} & \textbf{Cols Dropped} \\
  \midrule
  California Schools  & 3  & 3{,}870  & 14 \\
  European Football   & 8  & 23{,}209 & 12 \\
  Formula One         & 14 & 36{,}736 & 13 \\
  Superhero           & 10 & 1{,}056  & 9  \\
  \bottomrule
  \end{tabular}
\end{table}

\paragraph{Metrics.}
We measure result quality at different granularities with Execution Accuracy (Eq.~\ref{eq:ex}), Tuple-level F1 (Eq.~\ref{eq:tuplef1}), and Cell-level F1 (Eq.~\ref{eq:cellf1}), defined in Section~\ref{subsec:reward}. We also report the workflow efficiency in terms of wall-clock latency and LLM-operator costs.
\revise{For SemBench, we report its official aggregate Quality metric~\cite{lao2025sembench}.}

\subsection{Implementation Details}
\label{subsec:impl-details}

\paragraph{Testbed settings.}
All experiments run on an Ubuntu 22.04 server equipped with two Intel Xeon Platinum 8352Y CPUs (64 cores / 128 threads) and 504\,GB RAM. No GPU is required because all LLM inference is served through cloud API endpoints. We use DuckDB v1.5.0 as the query execution engine, exposed to the agent via a Model Context Protocol (MCP) stdio server.

\paragraph{Systems.}
On the same 80-query SWAN evaluation split, we compare \revise{all five systems}:
\begin{itemize}[leftmargin=*]
  \item \revise{\textbf{LOTUS~\cite{patel2025semanticoptimization}.} A semantic-query engine that executes declarative programs. For SWAN, a planner first uses the benchmark-provided gold SQL and BlendSQL annotations to translate the natural-language query into a fixed LOTUS program, which LOTUS executes without gold-output access or runtime replanning.}
  \item \revise{\textbf{Palimpzest~\cite{palimpzest2025}.} A declarative AI data-processing engine that likewise consumes an executable program. We use the same two-stage oracle-assisted translation and execution protocol with its native cost--quality optimizer.}
  \item \textbf{Agentic Text2SQL.} An LLM planner iteratively constructs and refines SQL, but query execution invokes no LLM operator. \revise{The planner may still encode parametric knowledge as SQL literals or deterministic rules. This control therefore isolates execution-time semantic operators rather than all LLM-supplied knowledge.}
  \item \textbf{Agentic BlendSQL.} The agent iteratively refines and executes hybrid queries through BlendSQL engine (v0.1.14), which embeds LLM operators as inline ingredients within SQL rather than exposing them as separate tool calls.
  \item \textbf{Agentic Query Execution (ours).} The agent receives the full MCP tool suite with separate semantic operators. We report results both without and with EnumGRPO. The latter augments the agent with an experience pool of distilled planning heuristics.
\end{itemize}

\revise{\paragraph{Agent settings.}
The agent is built on the OpenHands SDK v1.11.5. In the primary SWAN comparison, all five systems use the same model family. Agent planning and fixed-program compilation for LOTUS and Palimpzest use Claude Sonnet 4.6 (\$3 input, \$0.30 cache read, and \$15 output per 1M tokens). LLM operators in our system and semantic operators in BlendSQL, LOTUS, and Palimpzest all use Claude Haiku 4.5 (\$1/\$0.10/\$5). For the primary comparison, all calls are routed through LiteLLM v1.81.16 to the same Claude endpoints. For the cross-backbone study, the planner and LLM operators use GPT-5.6 Sol (\$4/\$0.40/\$20) and GPT-5.6 Luna (\$0.20/\$0.02/\$1.20), respectively. Each query execution has a 30-minute wall-clock timeout. Agentic systems are additionally limited to 30 steps.}

\paragraph{EnumGRPO settings.}
Learning uses the 40-query split in batches of two, with five rollouts per query
and 10 concurrent executions per batch. The plan enumerator seeds each rollout
with a distinct plan-space assignment (Section~\ref{subsec:plan-space}). The
resulting 200 rollouts complete in roughly 8.4 wall-clock hours. The composite
reward weights are $w_{\textsc{ex}} = 0.5$, $w_{\textsc{tuple}} = 0.3$,
$w_{\textsc{cell}} = 0.2$, with cost penalty weight
$w_{\textsc{cost}} = 0.2$ (split as $w_{\textsc{token}} = 0.7$,
$w_{\textsc{step}} = 0.3$). \revise{We assign half of the quality weight to
exact correctness and retain tuple- and cell-level partial credit to reduce
reward sparsity. The smaller cost penalty primarily distinguishes similarly
accurate rollouts; its token/step split emphasizes direct LLM usage while still
discouraging long workflows. We fix all weights before learning and do not tune
them.}


\newcommand{\std}[1]{{\scriptsize$\pm$\hspace{0.5pt}#1}}

\raggedbottom
\begin{table*}[t]
    \centering
    \caption{\revise{Overall results (mean$\pm$SD over 3 runs). Panel~A compares the primary systems on SWAN. LOTUS$^\dagger$ and Palimpzest$^\dagger$ additionally use SWAN-provided oracle information to assist program synthesis. Panel~B reports learning ablations and cross-database transfer. Panels~C and~D evaluate cross-backbone transfer and pure-SQL workloads, respectively. LLM-op metrics are workload totals. A dash here denotes no LLM-op access, whereas 0 indicates that LLM-ops are available but not invoked.}}
    \label{tab:main-results}
    \resizebox{\textwidth}{!}{%
    \begin{tabular}{@{}lrrrrrr@{}}
    \toprule
    \textbf{System} & \textbf{EX (\%)$\uparrow$} & \textbf{Tuple-F1 (\%)$\uparrow$} & \textbf{Cell-F1 (\%)$\uparrow$} & \textbf{Avg. Time (s)$\downarrow$} & \textbf{LLM-op Total Tokens$\downarrow$} & \textbf{LLM-op Total Cost (\$)$\downarrow$} \\
    \midrule
    \multicolumn{7}{@{}l}{\textbf{\textit{A. Primary evaluation: SWAN, Claude backbone (80 queries)}}} \\
    LOTUS$^\dagger$ & 20.8\std{3.2} & 25.3\std{1.7} & 30.4\std{2.1} & 166\std{18} & 2{,}343{,}833\std{85{,}938} & 2.5969\std{0.0980} \\
    Palimpzest$^\dagger$ & 12.5\std{1.3} & 17.1\std{2.3} & 18.5\std{1.7} & 585\std{45} & 30{,}766{,}853\std{8{,}729{,}678} & 47.5588\std{13.4816} \\
    Agentic Text2SQL & 19.2\std{1.4} & 22.8\std{2.2} & 28.0\std{2.2} & \textbf{142\std{12}} & -- & -- \\
    Agentic BlendSQL & 30.0\std{0.0} & 34.6\std{0.9} & 37.4\std{0.9} & 346\std{53} & 81{,}149{,}547\std{10{,}362{,}371} & 273.9520\std{34.3680} \\
    Agentic Query Execution \\
    \quad w/o experience pool & 32.1\std{1.9} & 36.5\std{2.1} & 39.9\std{1.3} & 235\std{62} & 2{,}646{,}347\std{206{,}756} & 3.1120\std{0.2880} \\
    \textbf{\quad w/ EnumGRPO (full method)} & \textbf{35.4\std{0.7}} & \textbf{38.1\std{0.8}} & \textbf{41.4\std{1.2}} & 153\std{14} & \textbf{739{,}600\std{266{,}288}} & \textbf{0.8640\std{0.3120}} \\
    \midrule
    \multicolumn{7}{@{}l}{\textbf{\textit{B. Learning ablations and cross-database transfer: SWAN, Claude backbone (80 queries)}}} \\
    Agentic Query Execution \\
    \quad EnumGRPO w/o plan enumeration & 26.7\std{0.7} & 29.2\std{0.8} & 32.9\std{1.2} & 125\std{4} & 284{,}694\std{229{,}861} & 0.3293\std{0.2621} \\
    \quad EnumGRPO w/o grouped comparison & 27.9\std{1.4} & 30.4\std{2.7} & 33.3\std{3.6} & 92\std{3} & 169{,}444\std{114{,}530} & 0.2115\std{0.1282} \\
    \quad EnumGRPO (cross-database transfer) & 32.9\std{2.2} & 36.0\std{1.6} & 39.2\std{1.6} & 110\std{4} & 656{,}576\std{215{,}943} & 0.7683\std{0.2385} \\
    \midrule
    \multicolumn{7}{@{}l}{\textbf{\textit{C. Cross-backbone transfer: SWAN, GPT-5.6 backbone (80 queries)}}} \\
    Agentic Query Execution \\
    \quad w/o experience pool & 35.0\std{0.0} & 41.7\std{0.8} & 44.1\std{0.8} & 198\std{47} & 12{,}029{,}547\std{2{,}509{,}198} & 8.4969\std{1.7188} \\
    \quad w/ Claude-learned experience pool & 35.4\std{0.7} & 42.1\std{0.5} & 44.3\std{0.5} & 177\std{41} & 4{,}325{,}254\std{1{,}664{,}380} & 2.8425\std{1.2239} \\
    \midrule
    \multicolumn{7}{@{}l}{\textbf{\textit{D. Pure-SQL workloads, Claude backbone (Spider: \emph{extra-hard} split, 166 queries; BIRD: \emph{challenging} split, 231 queries)}}} \\
    Agentic Query Execution \\
    \quad Spider, w/o experience pool & 72.9\std{1.2} & 78.3\std{0.4} & 86.0\std{0.4} & 24.1\std{0.3} & 0 & 0 \\
    \quad Spider, w/ SWAN-learned pool & 73.9\std{0.7} & 78.9\std{0.2} & 86.7\std{0.3} & 23.0\std{0.4} & 0 & 0 \\
    \cmidrule(lr){1-7}
    \quad BIRD, w/o experience pool & 38.7\std{0.7} & 40.8\std{0.7} & 61.5\std{0.9} & 33.4\std{0.1} & 0 & 0 \\
    \quad BIRD, w/ SWAN-learned pool & 39.3\std{0.9} & 42.0\std{0.4} & 62.7\std{0.6} & 28.3\std{0.5} & 0 & 0 \\
    \bottomrule
    \end{tabular}%
    }
\end{table*}

\section{Experimental Results}
\label{sec:results}

\subsection{Main Results}
\label{subsec:main-results}

\revise{Table~\ref{tab:main-results} reports our primary evaluation on SWAN. Agentic Query Execution with EnumGRPO achieves the highest execution accuracy while reducing total LLM-operator cost by 72\% relative to execution without the experience pool. The lower accuracy of Agentic Text2SQL shows the value of semantic operators. Although oracle-assisted LOTUS and Palimpzest receive additional benchmark information, both are less accurate and cost roughly 3$\times$ and 55$\times$ more, respectively. Agentic BlendSQL comes closest in accuracy but costs 317$\times$ more. Together, these results show the benefit of constructing semantic workflows adaptively from intermediate data.}

\revise{Trace analysis explains the savings. The learned experiences reduce LLM-operator-using queries from 51 to 40, and, among those queries, reduce average LLM-operator tokens from 52K to 18K. Output tokens likewise fall from 2,301 to 775. These shifts reflect greater use of SQL-visible evidence and code-driven rules, together with more targeted row scopes and narrower projections.}

\revise{The same mechanism explains the gaps to the semantic-system baselines. LOTUS and Palimpzest execute oracle-assisted programs fixed before query execution, so their semantic work cannot adapt to intermediate results. BlendSQL pushes SQL predicates before LLM ingredients when possible~\cite{zhao2025hybrid}, but each ingredient remains at its declared query position, leading to 1.01M LLM-operator tokens per query on SWAN. Agentic Query Execution instead controls when LLM operators run and how much data they receive.}
\vspace{5em}
\subsection{\revise{Learning Ablations}}
\label{subsec:ablations}

\revise{EnumGRPO combines structured plan enumeration for rollout diversity with GRPO-style grouped comparison for experience distillation. We conduct two ablations using the same five rollouts per query, reward, and pool-update procedure. Without enumeration, vanilla training-free GRPO \cite{cai2025trainingfree} retains grouped distillation but samples rollouts by temperature. It cuts LLM-operator cost by 62\% but sacrifices accuracy, suggesting that undirected sampling underexplores structurally distinct, higher-quality workflows. Without grouped comparison, the ungrouped variant retains enumeration but distills each trajectory independently like in Reflexion \cite{DBLP:conf/nips/ShinnCGNY23}. It lowers LLM-operator use but substantially degrades accuracy, suggesting that independent reflection overgeneralizes cheap SQL-first behavior. Same-query comparison improves credit assignment by separating efficient plans from cheap but inaccurate ones.}

\subsection{\texorpdfstring{\revise{Transferability Analysis}}{Transferability Analysis}}
\label{subsec:transfer}

\revise{\paragraph{Cross-database transfer.}
To assess cross-database transfer, we evaluate each target database using an experience pool constructed only from the other three SWAN databases. It reaches 32.9\% EX at \$0.010/query, slightly above Agentic Query Execution without the experience pool while halving latency. These results show that the experiences transfer across unseen schemas, while target-workload guidance remains useful for peak accuracy (35.4\%).}

\revise{\paragraph{Cross-backbone transfer.}
To test backbone portability, we apply the frozen Claude-learned pool without additional learning to an independent GPT-5.6 planner/operator stack. It preserves EX (35.4\% versus 35.0\%) while reducing mean workflow length from 14.6 to 12.5, latency from 198s to 177s, and LLM-operator cost from \$0.106 to \$0.036 per query. Thus, storing the learned policy in the prompt enables cross-model transfer without retraining.}

\begin{table}[t]
  \centering
  \small
  \setlength{\tabcolsep}{2pt}
  \caption{\revise{SemBench \emph{Movie} results (mean$\pm$SD over three runs).
  LLM-operator costs are totals over 10 queries.
  $^\dagger$ denotes oracle programs provided by SemBench~\cite{lao2025sembench} and BlendSQL~\cite{blendsqlnew}.}}
  \label{tab:sembench}
  \begin{tabular}{@{}lrrr@{}}
  \toprule
  \textbf{System} & \textbf{Quality$\uparrow$} & \textbf{Avg. Time (s)$\downarrow$} & \textbf{LLM-op Cost (\$)$\downarrow$} \\
  \midrule
  LOTUS$^\dagger$ & .629\std{.031} & 558.5\std{69.2} & 86.430\std{6.390} \\
  Palimpzest$^\dagger$ & \textbf{.762\std{.004}} & 313.5\std{12.0} & 135.792\std{.027} \\
  BlendSQL$^\dagger$ & .712\std{.010} & 100.6\std{1.3} & 15.049\std{.005} \\
  Agentic Text2SQL & .458\std{.146} & 187.3\std{27.9} & -- \\
  \multicolumn{4}{@{}l}{Agentic Query Execution} \\
  \quad w/o experience pool & .684\std{.053} & \textbf{80.6\std{8.1}} & .720\std{.053} \\
  \quad w/ SWAN-learned pool & .713\std{.027} & 82.5\std{5.6} & \textbf{.645\std{.029}} \\
  \bottomrule
  \end{tabular}
\end{table}

\revise{\paragraph{Cross-workload semantic transfer.}
Table~\ref{tab:sembench} reports transfer to the complete SemBench \emph{Movie}
scenario. For LOTUS, Palimpzest, and BlendSQL, we use pre-written oracle programs as provided by the prior works~\cite{lao2025sembench,blendsqlnew}.
All semantic operators use the same Claude Haiku 4.5 model as ours. Without learning from any \emph{Movie} query, our system with the transferred experience pool
matches oracle-program BlendSQL in Quality at 23$\times$ lower LLM-operator cost and exceeds LOTUS at 134$\times$ lower cost. Relative to Palimpzest, it achieves a slightly lower Quality score at
210$\times$ lower cost. This indicates that the learned guidance
transfers to a distinct semantic workload while preserving a favorable
quality--cost tradeoff.}

\revise{\paragraph{Cross-workload pure-SQL transfer.}
To assess how well the learned heuristics transfer to pure-SQL workloads, we
apply the frozen SWAN pool without additional learning to
Spider~\cite{yu2018spider} and BIRD~\cite{li2024bird}. The pool preserves
execution accuracy on both datasets, and across all runs the agent invokes no
LLM data operators. It also shortens the average trajectory from 3.99 to 3.55
rounds on Spider and from 6.93 to 5.13 rounds on BIRD, reducing mean wall time
on both workloads by up to 15.3\% on BIRD. These results show that the
transferred experiences improve SQL planning efficiency without introducing
unnecessary semantic processing.} 
\vspace{5em}
\subsection{\revise{Failure Analysis}}
\label{subsec:failure-analysis}

\revise{We perform a trace-based analysis of one representative 80-query
EnumGRPO run. The 52 workflows that fail EX fall into four mutually exclusive
categories. Incorrect \texttt{NULL} values enter one of the two value-error
categories according to the producing mechanism.}
\begin{itemize}[leftmargin=*]
  \item \revise{\emph{Execution/output}---4 cases (7.7\%): timeout,
  unrecovered SQL/tool error, abnormal termination, or missing/unreadable
  output.}
  \item \revise{\emph{Row-set construction}---16 failures (30.8\%): the
  completed output contains missing, extra, duplicated, or incorrectly selected
  rows, including an empty result when the reference is nonempty.}
  \item \revise{\emph{Semantic-operator value}---17 failures (32.7\%): row
  identity and cardinality agree with the reference, but values produced using
  an LLM data operator are incorrect.}
  \item \revise{\emph{SQL/rule-based value}---15 failures (28.8\%): row
  identity and cardinality agree with the reference, but values produced by
  ordinary SQL, a synthesized deterministic rule, or a proxy are incorrect.}
\end{itemize}
\revise{Of the 52 failures, 48 involve row-set construction or value
generation, suggesting that the main bottleneck lies in constructing and
interpreting intermediate results. The near-even split among row-set
construction, semantic-operator values, and SQL/rule-based values shows that
reliable execution depends jointly on selecting the correct rows and
maintaining value correctness across semantic and deterministic stages. This
challenge is particularly acute on SWAN, where every query requires information
absent from relational data. Even with an appropriate workflow plan, an LLM
operator may produce a locally incorrect value. Because EX credits only a fully
correct result, such an error makes the entire query incorrect, whereas
Tuple-F1 and Cell-F1 expose the partial correctness that EX masks. These
finer-grained metrics are available during learning and evaluation, but deployment provides
no reference answer for localizing errors within a workflow. Fine-grained
verification of intermediate row sets and generated values therefore remains
the key challenge for improving reliability.}


\begin{table*}[t]
  \centering
  \caption{\revise{Manual categorization} of the 32 experiences across the five plan-space axes, with one example experience per axis from the final pool $\varepsilon$. The \emph{Polarity} column shows how many experiences favor each axis value.}
  \label{tab:exp-axes}
  \begin{tabular}{@{}lrrlp{9.2cm}@{}}
  \toprule
  \textbf{Axis} & \textbf{\#} & \textbf{\%} & \textbf{Polarity} & \textbf{Example Experience} \\
  \midrule
  Execution paradigm  & 9  & 28\% & code 7\;:\;data 2
    & \emph{``Use CDSCode prefix extraction and county-code mapping in SQL first for geographic resolution; skip LLM operators entirely, eliminating token cost.''} \\[3pt]
  Selectivity scope   & 8  & 25\% & targeted 5\;:\;full 3
    & \emph{``When ranking by a rate metric, never filter out NULL/zero numerators---they represent the lowest valid rates and must be included.''} \\[3pt]
  Operator placement  & 6  & 19\% & pre 3\;:\;post 3
    & \emph{``Never apply \texttt{DISTINCT} or \texttt{GROUP BY} before computing per-record values; collapse rows only after all row-level calculations are complete.''} \\[3pt]
  Projection width    & 6  & 19\% & narrow 3\;:\;wide 3
    & \emph{``When asked where to find information about an entity, run \texttt{SELECT *} on that entity's row to check for URL/reference columns before joining tables.''} \\[3pt]
  Operator type       & 3  &  9\% & aggregate 3\;:\;scalar 0
    & \emph{``Use \texttt{llm\_reduce} when related rows benefit from joint reasoning rather than independent per-row inference.''} \\[3pt]
  \bottomrule
  \end{tabular}
\end{table*}

\begin{figure*}[t]
  \centering  \includegraphics[width=\textwidth]{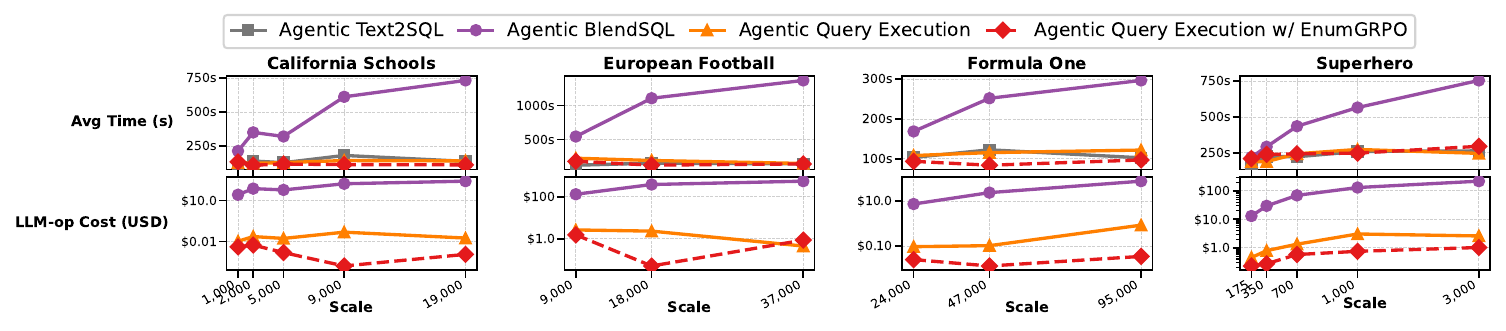}
  \caption{Scalability of average per-query latency (up) and total LLM-operator cost (down) as the primary table size in each database increases. To preserve the answer-bearing rows, \texttt{European Football} and \texttt{Formula One} cannot be downscaled below their original primary entity table sizes. Overall, Agentic BlendSQL’s latency and cost increase with scale, whereas both Agentic Query Execution configurations remain nearly flat by invoking LLM operators selectively on scoped intermediate results.}
  \label{fig:scalability}
\end{figure*}

\subsection{Scalability Analysis}
\label{subsec:scalability}

To test how latency and cost respond to data volume, we synthesize scaled copies of each database at $0.25\times$, $0.5\times$, $2\times$, and $4\times$ the primary entity table size and rerun the evaluation at each scale. Scaling uses one primary entity table per database as the cardinality knob, with non-primary tables left intact to preserve join validity. When downscaling, rows needed to preserve gold-answer values are protected before subsampling. Removing them would collapse the semantic gap that necessitates LLM inference, making queries purely relational or unanswerable.

The results are shown in Figure~\ref{fig:scalability}. Agentic BlendSQL diverges from the other systems as scale increases, most clearly on Superhero: from $0.25\times$ to $4\times$, its average wall time rises from 220s to 755s and its total LLM-operator cost rises from \$11.4 to \$164.9. This growth occurs because inline LLM ingredients process rows in intermediate relations whose cardinalities increase with data size. By contrast, Agentic Text2SQL and both Agentic Query Execution configurations remain nearly flat on both metrics, confirming that decoupling LLM operators from the query execution path is essential for predictable scaling.

A subtler pattern appears for Agentic Query Execution at larger scales. On California Schools and European Football, LLM-operator cost decreases rather than grows. The agent increasingly selects SQL-first or highly selective execution paths, such as pure SQL fallbacks or aggressive \texttt{WHERE IN} pre-filters. This behavior shows that agentic execution can adaptively bound semantic-operator cost as data size increases. It also highlights a tradeoff: overly conservative routing may skip useful semantic evidence. EnumGRPO shows the same tendency while keeping costs lower overall (Superhero $0.25\times$: \$0.18 to $4\times$: \$0.81). Future optimizers could improve scale robustness by conditioning learned experiences on relation cardinality and expected semantic coverage.

\subsection{Analysis of the Learned Experience Pool}
\label{subsec:experience-analysis}

\begin{figure}[t]
  \centering
  \includegraphics[width=0.9\columnwidth]{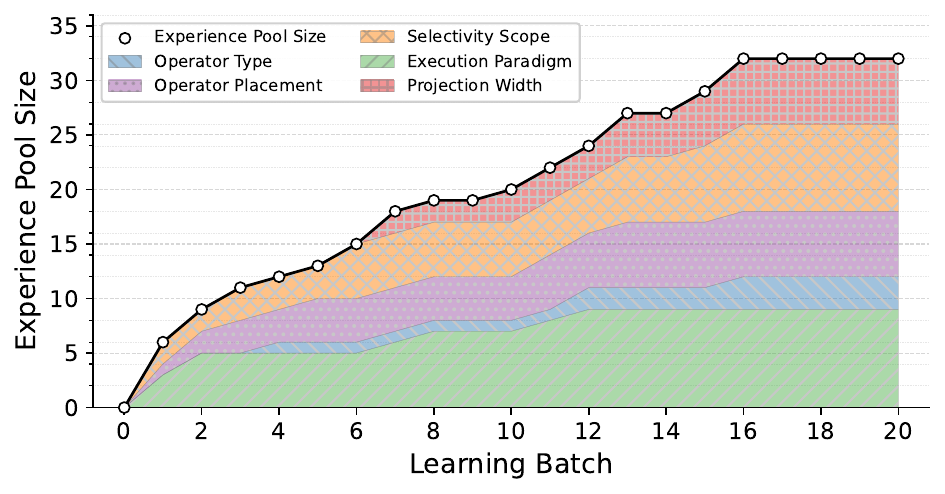}
  \caption{Experience-pool growth by axis, showing rapid early accumulation and convergence at later batches.}
  \label{fig:exp-growth}
\end{figure}

To understand what the optimizer learns, we examine how the experience pool
$\varepsilon$ grows during learning (Figure~\ref{fig:exp-growth}) and analyze
its final 32 entries by plan-space axis (Table~\ref{tab:exp-axes}). We form
these categories manually by inspecting each experience and assigning it to the
single axis most directly addressed by its recommendation. The table reports
the resulting distribution and polarity and provides one representative
example per axis from the pool.

The pool grows quickly in early batches and plateaus at 32 around batch~16, suggesting that later rollouts mainly reinforce learned heuristics. Execution-paradigm experiences appear first and remain the largest category, consistent with code-driven vs.\ data-driven execution being the dominant cost lever. Projection-width experiences emerge later, suggesting that context-width decisions become useful after broader execution and scoping heuristics have been learned.

The final pool is dominated by \emph{execution paradigm} and \emph{selectivity scope}, which together account for over half of all entries. This concentration matches the steepest part of the quality-cost tradeoff: execution paradigm determines whether LLM usage is $O(1)$ or $O(|R|)$, while selectivity scope controls how many rows reach those calls. By contrast, the small number of \emph{operator type} experiences suggests that scalar-vs.-aggregate choices are often subsumed by execution paradigm, since code-driven workflows avoid per-row LLM operators entirely. The remaining decisions are more conditional: \emph{operator placement} and \emph{projection width} are balanced across the polarities, indicating that the optimizer learns situational rules rather than blanket preferences.

\subsection{Learning Cost Analysis}
\label{subsec:tradeoff}
\revise{EnumGRPO incurs a material but one-time learning overhead. Its 200 rollouts over 40 labeled queries take 8.4 hours under ten-way rollout concurrency. Recorded calls total \$77.32: \$76.57 for rollout planning, \$0.40 for context compaction, and \$0.35 for semantic operators. Deployment freezes the resulting pool and performs no enumeration or policy update. Using this recorded learning cost, the representative original-scale ($1\times$) run implies a break-even of approximately 1{,}677 queries from an end-to-end planner-plus-operator saving of \$0.046/query. The threshold is workload-dependent. When primary-table cardinality doubles ($2\times$), the larger \$0.137/query saving lowers the threshold to approximately 564 queries. Together with Figure~\ref{fig:scalability}, these results show that EnumGRPO suits recurring workloads in which the experience pool can be reused. For one-off or low-volume workloads below these break-even points, the learning cost is not amortized, and Agentic Query Execution without the experience pool is more economical.}


\section{Related Work}
\subsection{Semantic Query Processing}
Recent systems have shown how to integrate semantic inference into data processing engines. BlendSQL extends SQL with LLM-backed operators such as \texttt{LLMMap} and \texttt{LLMQA} for hybrid question answering \cite{blendsql2024}. Zhao~\emph{et~al.} study hybrid querying over relational databases and LLMs, introducing the SWAN benchmark and the HQDL baseline based on schema expansion and UDF-style calls \cite{zhao2025hybrid}. CAESURA treats language models as planners that compile natural-language requests into multi-modal execution pipelines \cite{caesura2023}, while \emph{nsDB} argues for a neuro-symbolic DBMS with joint latency-accuracy objectives \cite{nsdb2024}. Palimpzest goes further by casting AI-powered analytics as declarative optimization over model, prompt, and execution choices along a Pareto frontier \cite{palimpzest2025}, and Weaver dynamically interleaves SQL and LLM reasoning for table question answering \cite{weaver2025}. These engines establish the importance of hybrid symbolic-neural analytics, but they largely optimize a fixed query, declarative program, or generated plan rather than deciding online from intermediate results which operator to apply next.

\subsection{Semantic Operator Optimization}
Another line of work focuses on optimizing the expensive semantic operators themselves. LOTUS introduces semantic operators with statistical accuracy guarantees and optimization strategies for filtering, joining, grouping, and top-$k$-style workloads \cite{patel2025semanticoptimization}. iPDB extends SQL with ML and LLM predicates and optimizes semantic queries via predicate ordering, prompt deduplication, and multi-row marshaling \cite{kumarasinghe2026ipdb}. Outside the database engine proper, Liu~\emph{et~al.} reorder LLM requests in relational workloads to improve cache locality and serving efficiency \cite{liu2024optimizing}. Classical work on efficient execution of UDFs \cite{foufoulas2023efficient} also provides a useful structural analogue, since it likewise shows that operator ordering and implementation details can dominate runtime. Our setting differs because the optimization target is not a single semantic operator or a fixed semantic SQL plan, but an execution-time workflow whose SQL and LLM steps, placement, and granularity are chosen jointly as execution unfolds.

\subsection{Learned Query Optimization}
Adaptive query optimization and learned query optimizers provide the database lineage for our approach. Eddies pioneered continuously adaptive query processing \cite{avnur2000eddies}. Bao learns per-query hint policies on top of an existing optimizer \cite{bao2021}, Balsa learns an optimizer without expert demonstrations via reinforcement learning \cite{balsa2022}, and Lero uses learning-to-rank to improve plan selection in existing DBMSs \cite{lero2023}. SkinnerDB is especially relevant because it learns join orders during execution and frames the problem through regret-bounded processing \cite{skinnerdb2021}. Learned cost models further inform this space by improving plan assessment for complex workloads \cite{li2024learned}. Our work extends this adaptive optimization perspective from conventional relational operators to heterogeneous SQL and semantic operators, where both execution cost and answer quality are data-dependent and may only become clear after intermediate results are observed.

\section{Conclusion}
\label{sec:conclusion}

\revise{We formulate agent workflow optimization, where planning interleaves with execution over intermediate results. EnumGRPO enumerates workflows over five axes and distills quality--cost feedback into reusable heuristics. On SWAN, it achieves 35.4\% execution accuracy at \$0.011 per query in LLM-operator cost. Comparisons with existing hybrid query systems including LOTUS, Palimpzest, and BlendSQL, plus cross-backbone and cross-workload evaluations, provide evidence of transfer across schemas, workloads, and model families. Future directions include task-conditioned experience retrieval, continual learning on evolving workloads, and stronger validation of intermediate workflow results. Beyond EnumGRPO, our findings point to a broader principle: as database systems incorporate approximate, expensive semantic operators, optimization must move beyond static plans toward learned policies that adapt to uncertainty, evidence, and cost at runtime.}

\bibliographystyle{ACM-Reference-Format}
\bibliography{references}

\end{document}